# Efficient Multidimensional Regularization for Volterra Series Estimation

Georgios Birpoutsoukis, Péter Zoltán Csurcsia, Johan Schoukens

*Vrije Universiteit Brussel, Department of Fundamental Electricity and Instrumentation,
Pleinlaan 2, B-1050 Elsene, Belgium (e-mail: georgios.birpoutsoukis@vub.ac.be)*

Abstract: **This paper presents an efficient nonparametric time domain nonlinear system identification method. It is shown how truncated Volterra series models can be efficiently estimated without the need of long, transient-free measurements. The method is a novel extension of the regularization methods that have been developed for impulse response estimates of linear time invariant systems. To avoid the excessive memory needs in case of long measurements or large number of estimated parameters, a practical gradient-based estimation method is also provided, leading to the same numerical results as the proposed Volterra estimation method. Moreover, the transient effects in the simulated output are removed by a special regularization method based on the novel ideas of transient removal for Linear Time-Varying (LTV) systems. Combining the proposed methodologies, the nonparametric Volterra models of the cascaded water tanks benchmark are presented in this paper. The results for different scenarios varying from a simple Finite Impulse Response (FIR) model to a 3rd degree Volterra series with and without transient removal are compared and studied. It is clear that the obtained models capture the system dynamics when tested on a validation dataset, and their performance is comparable with the white-box (physical) models.**

1. Introduction

In the field of system identification, modelling of nonlinear systems is one of the most challenging tasks. One possibility to model nonlinear dynamics in a nonparametric way is by means of the Volterra series ( [1]). The use of the series can be quite beneficial when precise knowledge about the exact nature of the nonlinear system behaviour is absent. Moreover, it has been shown in [2] that a truncated version of the infinite Volterra series can approximate the output of any nonlinear system up to any desired accuracy as long as the system is of fading memory. Practically speaking, it is sufficient that the influence of past input signals to the current system output decreases with time, a property that is met quite often in real systems, such as the cascaded water tanks system considered in this paper.

However, it is quite often possible that an accurate nonparametric fitting of nonlinear dynamics requires an excessive number of model parameters. Under this condition, either long data records should be available, resulting in computationally heavy optimization problems and long measurements, or the estimated model parameters will suffer from high variance. This is the reason why the series has in general been of limited use (for echo cancellation in acoustics [3], [4] and for physiological systems [5]), and mostly in cases where short memory lengths and/or low dimensional series for the modelling process were sufficient (e.g. [6]). In the field of mechanical engineering numerous applications of the Volterra series can be found (e.g. for mechanical system modelling [7], [8], [9], for damage detection [10], [11]). We refer to [12] for an extended survey on the Volterra series and its engineering applications.

In this paper, we present a method to estimate efficiently finite Volterra kernels in the time domain without the need of long measurements. There are several studies available in the literature where methods are proposed to reduce the dimensionality (memory) issues of the Volterra series estimation problem, for example, with the use of orthonormal basis functions [13], [14] or tensor decompositions [15]. The main disadvantage of these techniques lies in the choice of the poles of the basis functions. The latter issue is even more involved in the case of expanding a Volterra series model with basis functions [14].

The method presented in this work is based on the regularization methods that have been developed for FIR modelling of Linear Time-Invariant (LTI) systems [16], while results exist also for the case of Frequency Response Function (FRF) estimation [17]. In the aforementioned studies, the impulse response coefficients for a LTI system are estimated in an output error setting using prior information during the identification step in a Bayesian framework. The knowledge available a priori for the FIR coefficients was related to the fact that the Impulse Response Function (IRF) of a stable LTI system is exponentially decaying, and moreover, there is a certain level of correlation between the impulse coefficients (smoothness of estimated response).

The regularization methods introduced for FIR modelling are extended to the case of Volterra kernels estimation using the method proposed in [18]. The benefit of regularization in this case with respect to FIR modelling is even more evident given the larger number of parameters usually involved in the Volterra series. Prior information about the Volterra kernels includes the decaying of the kernels as well as the correlation between the coefficients in multiple dimensions. Due to the fact that in case of long measurements and higher order Volterra series, the requested memory needs can be more demanding than the available resources, a memory and computational complexity saving algorithm is provided as well.

In this work, the regularized Volterra kernel estimation technique is combined with a method for transient elimination which plays a key role in this particular benchmark problem because each measurement contains transient. Due to the fact that the measurement length is comparable to the number of parameters, it is necessary to eliminate the undesired effects of the transient as much as possible. The proposed elimination technique uses a special LTI regularization method based on the ideas of an earlier work on nonparametric modelling of LTV systems [19]. It is important to highlight that transient elimination has been not applied yet for the case of Volterra series estimation.

The paper is organized as follows: Section 2 introduces the regularized Volterra kernel estimation method. Section 3 deals with the excessive memory needs of long measurements and large number of parameters. In Section 4 the proposed method for the transient removal is presented. In Section 5 the benchmark problem is formulated and the concrete benchmark results are shown illustrating the efficiency of the combination of the two proposed methods for modelling of the cascaded water tanks system. Early results on the cascaded water tanks benchmark problem can be found in [20]. Finally, the conclusions are provided in Section 6.

2. The nonparametric identification method

*2.1 The model structure*

It is assumed that the true underlying nonlinear system can be described by the following truncated discrete-time Volterra series [1]:

$$y_{meas}(n) = h_0 + \sum_{m=1}^{M}(\sum_{\tau_1=0}^{n_m-1} \ldots \sum_{\tau_m=0}^{n_m-1} h_m(\tau_1, \ldots, \tau_m) \prod_{\tau=\tau_1}^{\tau_m} u(n-\tau)) + e(n) \qquad (1)$$

where $u(n)$ denotes the input, $y_{meas}(n)$ represents the measured output signal, $e(n)$ is zero mean i.i.d. white noise with finite variance $\sigma^2$, $h_m(\tau_1, \ldots, \tau_m)$ is the Volterra kernel of order $m = 1, \ldots, M$, $\tau_i, i = 1, \ldots, m$ denote the lag variables and $n_m - 1$ corresponds to the memory of $h_m$. The Volterra kernels are considered to be symmetric, which means that [12]:

$$h_n(\tau_1, \tau_2, \ldots, \tau_n) = h_n(\tau_2, \tau_1, \ldots, \tau_n) = \cdots = h_n(\tau_{i_1}, \tau_{i_2}, \ldots, \tau_{i_n}), \quad i_j \neq i_k \qquad (2)$$

$$i_1, i_2, \ldots, i_n \in (1, 2, \ldots, n), \quad j, k \in (1, 2, \ldots, n)$$

Due to symmetry, it can be easily shown that the number of coefficients to be estimated for a symmetric Volterra kernel of order $m \geq 1$ is $n_{h_m} = (\frac{1}{m!}) \prod_{i=0}^{m-1}(n_m - i)$. It is also important to clarify the difference between order and degree of the Volterra series with an example: the third degree Volterra series contains the Volterra kernels of order 0, 1, 2 and 3.

*2.2 The cost function*

Given $N$ input-output measurements from the system to-be-identified, equation (1) can be rewritten into a vectorial form as $Y_{meas} = K\theta + E$, where $\theta \in \mathbb{R}^{n_\theta}$, $n_\theta = 1 + \sum_{m=1}^{M} n_{h_m}$, contains vectorised versions of the Volterra kernels $h_m$, $K \in \mathbb{R}^{N \times n_\theta}$ is the observation matrix (the regressor), $Y_{meas} \in \mathbb{R}^N$ contains the measured output and $E \in \mathbb{R}^N$ contains the measurement noise (observation error).

In this work, the Volterra kernel coefficients in $\theta$ are estimated by minimizing the following regularized least squares cost function $V$ as a combination of the ordinary least squares cost function ($V_{LS}$) and the regularization term ($V_r$):

$$V = V_{LS} + V_r = E^T E + V_r = ||Y_{meas} - K\theta||_2^2 + \theta^T D \theta \qquad (3)$$

where the block-diagonal matrix $D \in \mathbb{R}^{n_\theta \times n_\theta}$ contains $(M + 1)$ submatrices penalizing the coefficients of the Volterra kernels. Given an appropriately structured penalizing matrix $D$ (see Section 2.3) and the cost function $V$ (3), the regularized least squares solution is given by:

$$\hat{\theta}_{reg} = arg \min_{\theta} V = (K^T K + D)^{-1} K^T Y_{meas} \qquad (4)$$

It can be observed that the Maximum Likelihood (ML) estimation of the Volterra coefficients can be computed by setting $D = 0$. The ML estimates suffer very often from high variance in the case of Volterra series estimation due to the large number of parameters in the model (curse of dimensionality).

*2.3 Prior knowledge and regularization*

To overcome the ML estimation issues, a block-diagonal matrix $D$ is constructed such that prior information about the underlying dynamics of the true system is taken into account during the identification procedure. The regularization matrix $D$ is constructed using a Bayesian perspective as explained in [21], [22]. It is computed via the product of an inverse covariance matrix $P = \mathbb{E}[\theta\theta^T]$ ($\mathbb{E}$ denoting the mathematical expectation operator) and the noise variance $\sigma^2$ as follows:

$$D = \sigma^2 P^{-1} \tag{5}$$

where $P \in \mathbb{R}^{n_\theta \times n_\theta}$ is a block-diagonal covariance matrix, $P = blkdiag(P_0, P_1, P_2, ..., P_M)$ and blkdiag is a function whose output is a block diagonal matrix with the matrix arguments on its diagonal. Each covariance matrix on the diagonal corresponds to a Volterra kernel of different order, namely $P_n = \mathbb{E}[\theta_n \theta_n^T]$ where $\theta_n$ denotes the vectorised version of the Volterra kernel $h_n$.

The matrix $P$ introduces the prior knowledge into the cost function as an a priori correlation and scaling between the elements of the IRFs. With respect to the prior knowledge used in this work, it is assumed that the Volterra kernels used to describe the true system are 1) decaying and 2) smooth.

The first property refers to stability not only for the linear term $h_1(\tau_1)$ but also for the higher dimensional impulse responses (i.e. higher dimensional Volterra kernels). Practically speaking, it is assumed that $h_m(\tau_1, ..., \tau_m) \to 0$ for $\tau_1, ..., \tau_m \to \infty$.

The second property of smoothness is related to the correlation between the Volterra coefficients. For the discrete-time Volterra series used in this paper, a smooth estimated Volterra kernel means that there exists a certain level of correlation between neighbouring coefficients, which decreases the larger the distance between two Volterra coefficients.

### 2.3.1 Kernel-based regression and the covariance matrix $P_1$ for the first order Volterra kernel

The specific choice of the kernels and the resulting covariance matrix, used to impose smoothness and exponential decaying of the impulse responses, have a major effect on the quality of the estimated model.

Note, that the term "kernel" refers to a covariance matrix and is not to be confused with a Volterra kernel. Throughout the paper the word "Volterra" is always used to distinguish between regularization kernels (covariance matrix) and the Volterra kernels in the series (1).

#### 2.3.1.1 Diagonal Correlated (DC) Kernels

This kernel has the flexibility to tune independently the properties of smoothness and the exponential decay. The DC kernel function for the first order Volterra kernel ($P_1 = \mathbb{E}[h_1(\tau_1) h_1(\tau_1)^T]$) is defined as follows:

$$P_{1,DC}(i,j) = c_1 \rho^{|i-j|} \alpha^{(i+j)/2} \tag{6}$$

where $c_1 \geq 0, 0 \leq \alpha < 1, |\rho| \leq 1$. The correlation length between adjacent impulse response coefficients is quantified by $\rho$ (i.e. it controls the smoothness property), and $\alpha$ scales the exponential decaying.

Note that in many cases the formulation of the DC kernel differs from (6), like in [23] but the expressions are equivalent.

#### 2.3.1.2 Tuned-Correlated (TC) Kernels

While the DC kernel gives a nice flexibility to tune the model behaviour, unfortunately it can lead to a very high computational complexity (3 parameters to tune). In some cases TC kernels – as a special form of DC – can provide a good balance between the flexibility and the computational load. When, in the DC kernel structure, $\rho$ is set equal to $\sqrt{\alpha}$ then it leads to the TC form:

$$P_{1,TC}(i,j) = c_1 \alpha^{max(i,j)} \tag{7}$$

where $c_1 \geq, 0 \leq \alpha < 1$.

Note, that the TC kernels are widely used in different regularization toolboxes (e.g. *ident* in Matlab).

#### 2.3.1.3 Other kernels

Depending upon the prior knowledge, different covariance matrices can be used as well. This holds true also for the higher order Volterra kernels (see next). A detailed comparison can be found in [16], [24], [25].

### 2.3.2 The covariance matrix $P_2$ for the second order Volterra kernel

The two facts of decaying (stability) and smoothness (correlation) for the second order Volterra kernel are encoded into the matrix $D$ (see (2)) using the method described in [18]. It is an extension of the regularization methods for linear impulse response [26] using the Diagonal Correlated (DC) kernel (again not to be confused with a Volterra kernel) to higher dimensions such that prior information is imposed to a multi-dimensional impulse response.

The properties of decaying and smoothness for the 2$^{nd}$ order Volterra kernel are encoded into the matrix $P_2$. The $(i,j)$-element, which corresponds to $\mathbb{E}[\theta_{2,i}\, \theta_{2,j}]\,\forall i,j$ where $\theta_{2,i}$, $\theta_{2,j}$ denote two Volterra coefficients in $\theta_2$, is given by [18]:

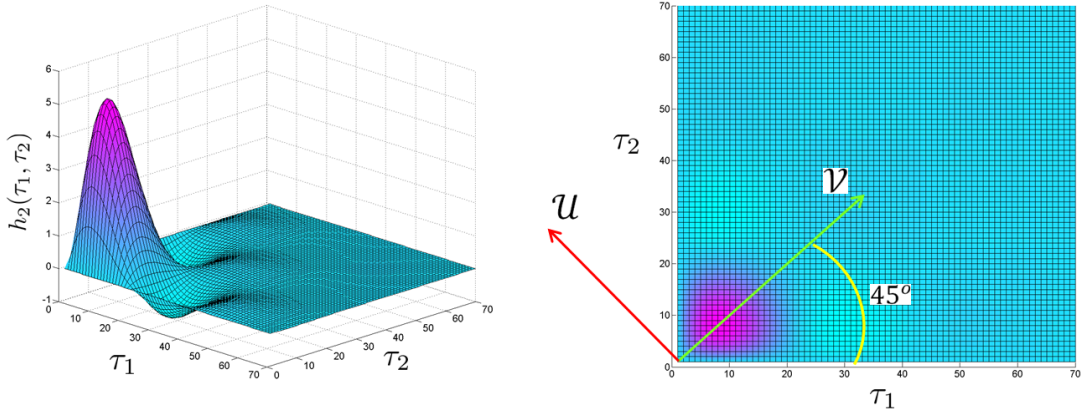

Fig 1. Left: An example of a smooth and exponentially decaying second order Volterra kernel. Right: X-Y view of the same Volterra kernel together with the two directions, along which prior information is imposed.

$$P_{2,DC}(i,j) = c_2\, \rho_\mathcal{U}^{||u_i|-|u_j||}\, \alpha_\mathcal{U}^{\frac{|u_i|+|u_j|}{2}}\, \rho_\mathcal{V}^{||v_i|-|v_j||}\, \alpha_\mathcal{V}^{\frac{|v_i|+|v_j|}{2}} \tag{8}$$

where the $\mathcal{U}, \mathcal{V}$ - coordinate system is rotated 45 degrees counter-clockwise with respect to the coordinate system $\tau_1$ and $\tau_2$ (see Fig.1, right):

$$\begin{bmatrix} \mathcal{U}_i \\ \mathcal{V}_i \end{bmatrix} = \begin{bmatrix} \cos(45^o) & -\sin(45^o) \\ \sin(45^o) & \cos(45^o) \end{bmatrix} \begin{bmatrix} \tau_{1,i} \\ \tau_{2,i} \end{bmatrix} \tag{9}$$

The so-called hyperparameters $\rho_\mathcal{U}$ and $\rho_\mathcal{V}$ are used to control the correlation between the coefficients along the $\mathcal{U}$ and $\mathcal{V}$ direction, respectively. The hyperparameters $\alpha_\mathcal{U}$ and $\alpha_\mathcal{V}$ determine the decay rate along the $\mathcal{U}$ and $\mathcal{V}$ direction, respectively. Hyperparameter $c_2$ is a scaling factor used to determine the optimal trade-off between the measured data and the prior information encoded in $P_2$ (the subscript 2 is used in $c_2$ to show that this hyperparameter is different from $c_1$ used in (6)–(7) for the first order Volterra kernel and in general different for every Volterra kernel in the series). Equation (8) can also be seen as the product of two DC kernels applied in the $\mathcal{U}$ and $\mathcal{V}$ directions, respectively. For a more detailed analysis on the proposed covariance matrix for the second order Volterra kernel, the interested reader is referred to [18].

### 2.3.3 Extension of the method to higher order Volterra kernel estimation

The Volterra kernel regularization method proposed for the second order Volterra kernel can be extended to cover the case of Volterra kernels of orders higher than two. In Fig. 2 a part of the third order kernel, described by the function $h_3(\tau_1, \tau_2, \tau_3)$ in the Volterra series (1), is depicted. Each sphere corresponds to a different value of the function $h_3$ (different point). The diagonal direction $\mathcal{V}$ is now given by the blue arrow with orientation (1,1,1) (in the $\tau_1, \tau_2, \tau_3$ coordinate system).

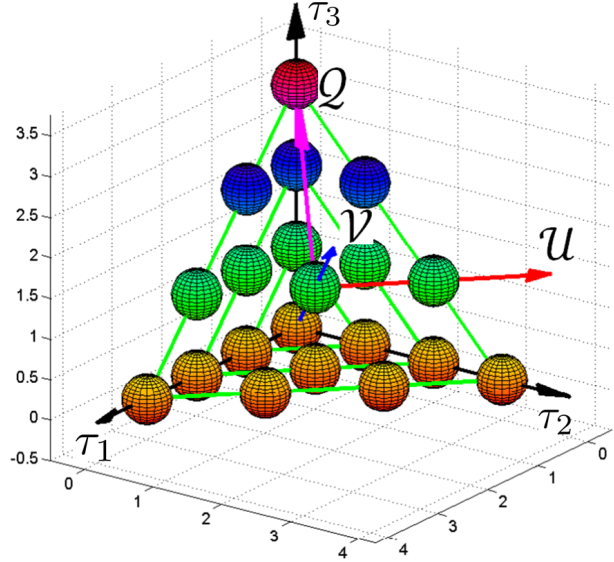

Fig 2. The three directions, along which prior information is imposed, for the third order Volterra kernel.

At each point on this direction, there corresponds a plane described by the equation $\tau_1 + \tau_2 + \tau_3 = constant$. In Fig. 2 these planes are delimited with green lines. Following the same reasoning as for the second order kernel, two extra directions are needed, in order to describe the prior knowledge of exponential decaying and smoothness on each of these planes. Therefore, the three directions $\mathcal{V}$ (orientation (1,1,1)), $\mathcal{U}$ (orientation (1,-1,0)) and $\mathcal{Q}$ (orientation (-1,-1,2)) can be used to formulate the covariance for the three dimensional impulse response, following the same reasoning as in (7). In this way, the concept of imposing prior information through regularization can be further extended to higher dimensional Volterra kernels.

## 2.4 Tuning of the hyperparameters

All the so-called hyperparameters, such as $c_1, \alpha, \rho$ in $P_1$, $c_2, \rho_\mathcal{U}, \rho_\mathcal{V}, \alpha_\mathcal{U}, \alpha_\mathcal{V}$ in $P_2$ as well as $\sigma^2$ (noise variance) and $P_0 = \mathbb{E}[h_0 \cdot h_0]$ are tuned with the use of the input and output data used for estimation. The nonparametric system identification method presented in this work consists essentially of two steps:

- Optimization of the hyperparameters to tune the matrix $P$ (and further $D$ in (5)).
- Computation of the model parameters (Volterra series coefficients) using (4).

The values of the hyperparameters can be tuned in many different ways. The theoretical aspects of the possible optimization techniques are beyond the scope of this paper. In this work the marginal likelihood approach [27], [28] and a residual analysis [29] are used because they need only one estimation (training) dataset for the entire estimation procedure. However, it is important to point out that a simple cross-validation technique [30] can be used as well, but it would not fall within the frames of the measurement benchmark problem since it would also require the use of the validation (testing) dataset [31]. Alternatively, the estimation dataset could be split into two parts (estimation and validation sets) such that cross validation could be used. Unfortunately, in this work it would be not an option since the estimation dataset of the benchmark problem is very short, and by splitting it into two would result in a situation where the number of parameters is much higher than the number of measurement samples available.

### 2.4.1 The marginal likelihood approach

The hyperparameters are optimized in this case by maximizing the marginal likelihood of the observed output, which boils down to:

$$\hat{\theta}_{hp} = \arg \min_{\theta_h} Y_{meas}{}^T \Sigma_Y^{-1} Y_{meas} + logdet\, \Sigma_Y \qquad (10)$$

where $\theta_{hp}$ is a vector containing all the hyperparameters and $\Sigma_Y = (KPK^T + \sigma^2 I) \in \mathbb{R}^{N \times N}$ represents the covariance matrix of the output vector $Y_{meas}$, and $I$ denotes the identity matrix. Practically speaking, the optimal set of hyperparameters given the measured output data determines the covariance matrix which is most likely to have generated the observed output.

The objective function (10) is non-convex in $\theta_{hp}$, therefore to minimize the risk of resulting in a local minimum, it is advised that multi-start optimization of the hyperparameters is performed. In this work, a nonlinear gradient based Matlab solver is used (*fmincon*) for the optimization of the marginal likelihood of the observed output (10).

The interested reader can refer to [32] for more techniques for the tuning of the hyperparameters.

*2.4.2 The residual analysis approach*

In case of residual analysis [29] the hyperparameters, the model and the measured outputs are analysed simultaneously. The output simulation is performed by using a set of (candidate) hyperparameters $\hat{\theta}_{hp}$ as $Y_{mod} = K\hat{\theta}_{reg}$, see (1),(4). In this terminology the residuals ($E$) are the differences between the simulated output and the measured output, i.e. $E = Y_{meas} - K\hat{\theta}_{reg}$.

It can be clearly seen that the residuals represent a part of the data that are not explained by the model. Ideally, if the system dynamics is well-modelled by $\hat{\theta}_{reg}$, then the residual term $E$ should only contain the noise samples. Therefore, the key idea in residual analysis is to analyze the statistical properties of $E$. In practice, this statistical analysis consists of two tests: the whiteness test and the independence test. If the residuals are white and independent of each other i.e. $E$ is the samples of the additive Gaussian noise, then an appropriate model order as well as set of hyperparameters has been retrieved.

The interested reader can refer to [29], [33], [34] for a detailed studies on the residual analysis.

*2.4.3 Computational concerns*

The inversion of $\Sigma_Y$ in (10) for the marginal likelihood maximization, which is equivalent to the one in (4) for the residual analysis approach, is performed multiple times during the optimization of the hyperparameters. This matrix inversion constitutes one of the most time-consuming parts of the nonparametric identification method presented in this paper due to its large size ($\Sigma_Y \in \mathbb{R}^{N \times N}$). In the following section, the problem of matrix inversion both at the level of the cost function as well as at the level of the hyperparameters optimization is tackled.

3. Coping with large datasets and excessive number of parameters

*3.1 Introduction*

In the past, engineers and scientists barely used Volterra models for nonparametric estimation due to the fact that it required 1) much more measurement samples than the number of model parameters, and 2) an excessive number of parameters. Injecting prior information, such as smoothness and decaying, decreases significantly the need for long data records. Further, there is a significant increase at the modelling quality due to the injection of prior knowledge, however, this is at the price of increased computational complexity and higher memory needs. The computational complexity increases with the degree of the Volterra series (and thus with the increasing number of parameters, see Fig. 3).

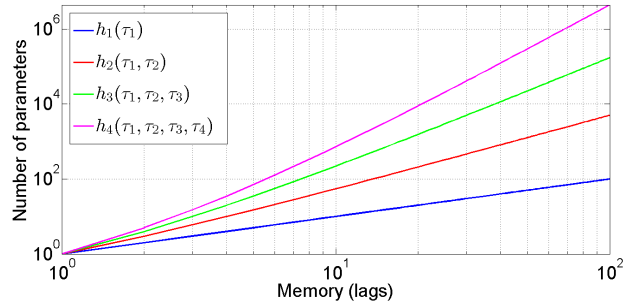

Fig 3. The number of model parameters up to the fourth order symmetric Volterra kernel is shown as a function of the memory length (truncation lag).

The exact computational time depends on many factors such as the model complexity, the way how the matrix algebra is implemented, the initial values of the hyperparameters, and the technique used to tune the hyperparameters. A significant amount of time can be saved by using different hyperparameter optimization strategies, and by choosing properly the initial values. The case of hyperparameter tuning using the Marginal Likelihood approach when the data length $N$ is much larger than the number of parameters $n_\theta$ has already been studied in [32]. However, this is not the case of the cascaded water benchmark problem (see Section 5.1). In general, the most time-consuming part is the matrix inversion in (4) which needs to be executed in each optimization cycle of the tuning of hyperparameters. Another (hidden) matrix inversion problem can be found in the penalization matrix $D \in \mathbb{R}^{n_\theta \times n_\theta}$. In general, the numerical values of $D$ are computed via scaled and inverted covariance matrices.

A further highly related issue is the exaggerated memory needs. In general, when LTI systems are identified by regularization technique, the memory needs are quite limited. However, considering the matrix sizes in this particular case, it can be clearly seen that by increasing the length of the measurement, the requested operational memory size grows linearly as $O(2N \times n_\theta + N)$, by increasing the number of parameters it grows quadratically as $O(3n_\theta \times n_\theta + n_\theta)$.

Further, the memory needs also grow as a power function of the number of lags to the Volterra degree i.e. $O(\sum_{m=1}^{M}(n_{h_m})^2)$, see Fig 3.

In order to illustrate the problem an example is given. Let us assume that a nonlinear system is modelled with the third degree Volterra model, the measurements contain 10 000 samples, the number of lags in the first and second order Volterra kernels is 100, the number of lags in the third order Volterra kernel is 50. In this case the requested memory size (with 8 byte floating precision) is more than 459 GB RAM. However, most of the personal computers do not have this capacity, therefore a solution is needed which would allow to run a higher degree Volterra kernel estimation on every (modern) computer.

*3.2 Proposed solution*

In this work a vectorised gradient based method is presented. By applying the proposed technique it is possible 1) to avoid the inversion in the regularised LS solution (4), 2) to avoid the inversion in the penalization matrices (5), and 3) to reduce the memory needs. It is important to remark that in [19] regularized two-dimensional IRFs were studied where one faced very similar memory issues presented here. The key idea of [19] is that instead of computing the 2D time-varying nonparametric model in one step for a large dataset, a special sliding window can be used for partial datasets resulting in a smaller computational load. The sliding window is applied over different, consecutive impulse response functions. Unfortunately, due to the problem formulation this technique cannot be applied here.

*3.3 Avoid inversion in the cost function*

At this point it is important to recall the original cost function (3). The exact solution of the regularized estimate – for a set of hyperparameters encoded in $D$– is given by (4). The inversion in (4) needs to be (re)calculated for each set of hyperparameters –in each hyperparameter optimization cycle. In order to avoid inversion of large matrices (i.e. in case of large $N$), it is advised to use an inversion-free gradient based method.

In order to obtain the proposed algorithm, first let us calculate the gradient of the cost function (3):

$$\frac{\partial}{\partial \theta} V = \frac{\partial}{\partial \theta}(V_{LS} + V_r) = \frac{\partial}{\partial \theta}(\|Y_{meas} - K\theta\|_2^2 + \theta^T D\theta) = \frac{\partial}{\partial \theta}(E^T E + \theta^T D\theta) = -K^T E + D\theta \qquad (11)$$

Next, let us introduce new terms for the proposed method. The Jacobian ($J \in \mathbb{R}^{n_\theta \times N+n_\theta}$) and gradient descent direction ($E_{gdd} \in \mathbb{R}^{N+n_\theta \times 1}$) matrices are defined as follows:

$$J = \frac{\partial}{\partial \theta}[E \quad \theta^T D] = [-K^T \quad D] \qquad (12)$$

$$E_{gdd} = [E \quad \theta]^T \qquad (13)$$

In the proposed gradient based method several iterations –i.e. updates– will be needed. The actual iteration number is indicated hereinafter in the subscript of the variables. The initial value of the parameter estimate vector ($\hat{\theta}_0$) is advised to be set to zero. At $(k+1)th$ step, the new value of the parameter estimate is given by:

$$\hat{\theta}_{k+1} = \hat{\theta}_k - \lambda J E_{gdd} \qquad (14)$$

where $\lambda \in \mathbb{R}$ is used as a weighting factor for the gradient descent direction.

Using the above-mentioned terms, for a given penalization matrix $D$, the algorithm of the inversion-free regularized solution is the following:

1. Initialize $\hat{\theta}_0$.
2. Calculate the initial cost function $V_0 = V(\hat{\theta}_0)$.
3. Repeat until the maximum number of iterations is reached:
    a) Compute the descent direction ($E_{gdd}$) and initialize $\lambda$.
    b) Calculate the new candidate value of the parameter estimate vector ($\hat{\theta}_{k+1}$).
    c) If $|V_{k+1} - V_k|$ is smaller than the tolerance level of the cost function change, then leave 3.
    d) If $V_{k+1} < V_k$ then update $\hat{\theta}_{k+1}, V_{k+1}$ and go back to 3.
    e) If $\lambda$ is lower than its threshold (minimum allowed value), then leave 3.
    f) Reduce the value of $\lambda$ and go back to 3.b.

For the implementation of the algorithm it is important to emphasize that the default value of $\lambda$ should be small ($\lambda \ll 1$), otherwise the step in the descent direction (i.e. the direction towards the minimum) will be too large and it might diverge. In the other hand, if $\lambda$ is too small (close to zero) then the necessary number of iterations will be excessive. In this work a simple line search algorithm is used: at each unsuccessful cost function evaluation (see 3.d), the value of $\lambda$ is decimated (see 3.f). After each new iteration (3.a) $\lambda$ is restored to its default value. 3.c and 3.e of the algorithm make it possible to stop the optimization procedure if the change in the cost function or in $\lambda$ is insignificantly too small (i.e. when it would make no significant gain in the modelling quality).

Of course, the above mentioned algorithm can be extended to any kind of gradient method. The presented algorithm is proposed because its computational resources are low and it is inversion free (compared, for instance, to Levenberg–Marquardt algorithm [35]).

A problem with this formulation is that, even though, we got rid of the explicit inversion in the cost function, there is still implicit inversion involved in the cost function. The numerical values of $D$ is still computed by matrix inversion (5). Next, an alternative possible way of constructing the penalization term is discussed.

*3.4 Avoid the inversion in the penalization matrices*

The penalization term $D$ is usually computed via the inverse covariance matrix (see Section 2.3). The covariance matrices are built by using so-called regularization kernel functions. [36] presents a possibility to directly inject prior information at the cost function level, instead of creating covariance matrices and inverting them. The key idea is to include the regularization term in the cost function as a FIR filtering operation on the parameters to be estimated (see (17)). The available prior knowledge about the system can be directly injected into $D$ instead of injecting in $P$ and then inverting it. In this way, instead of defining the problem based on the covariance matrix, one can focus directly on the cost function. The approach proposed starts from the following triangular matrix decomposition by $F \in \mathbb{R}^{n_\theta \times n_\theta}$:

$$D = \sigma^2 P^{-1} = \sigma^2 F^T F \tag{15}$$

The cost function is then reformulated as follows:

$$\hat{\theta}_{reg} = V_{LS} + \theta^T D \theta = V_{LS} + \sigma^2 \theta^T F^T F \theta = V_{LS} + \sigma^2 \|F\theta\|^2 \tag{16}$$

In the equation above, $F$ can be seen as a prefiltering FIR operator on the coefficients of the impulse response, before they enter the cost function and are penalized through the regularization term. This means that the regularization filter matrix $F$ should be defined in such a way that it incorporates the system properties one needs to penalize to obtain the desired model. In fact, the rows of $F$ are penalization FIR filters at different time instances ($n$):

$$F_1 = \begin{bmatrix} h_F(0)_{n=0} & h_F(1)_{n=0} & \dots & \dots & h_F(n_{m_1})_{n=0} \\ 0 & h_F(0)_{n=1} & h_F(1)_{n=1} & \dots & h_F(n_{m_1}-1)_{n=1} \\ 0 & 0 & \ddots & \ddots & \ddots \\ 0 & 0 & \dots & 0 & h_F(0)_{n=n_{m_1}-1} \end{bmatrix} \tag{17}$$

[36] proposes a unified filter approach for $F_1$: the $h_F(j)_{n=i}$ parameters are the decayed FIR coefficients of different low-pass/high-pass/band-pass filters which provide a great deal of flexibility. However, this unified filter approach is not recommended to be directly used here, since it would require a higher number of hyperparameters. In this work, instead of using the unified filter approach, DC and TC regularization filter functions are defined with the help of Cholesky decomposition [36], [37]. For the sake of simplicity, $c_1$ is set to 1. In this case the first order regularization filter functions are:

$$F_{1,DC}(i,j) = \begin{cases} \sqrt{\frac{1}{\alpha^i(1-\rho^2)}} & \text{if } i = j, \ i < n_\theta \\ \sqrt{\frac{1}{\alpha^i}} & \text{if } i = j = n_\theta \\ -\sqrt{\frac{1}{\alpha^i(1-\rho^2)}} & \text{if } i = j-1 \\ 0 & \text{otherwise} \end{cases} \tag{18}$$

$$F_{1,TC}(i,j) = \begin{cases} \sqrt{\frac{1}{\alpha^i(1-\alpha)}} & \text{if } i = j, \ i < n_\theta \\ \sqrt{\frac{1}{\alpha^i}} & \text{if } i = j = n_\theta \\ -\sqrt{\frac{1}{\alpha^i(1-\alpha)}} & \text{if } i = j-1 \\ 0 & \text{otherwise} \end{cases} \tag{19}$$

The hyperparameters above ($c_1$, $\alpha$, $\rho$) are perfectly identical to the previously used hyperparameters (see Section 2.3.1).

Note, that higher order filter functions ($F_{2,DC}$, $F_{3,DC}$, etc.) can be defined by the combination of first order filter functions as it is proposed in Section 2.3.2-2.3.3.

In this special case, the Jacobian and gradient descent direction matrices are redefined as follows:

$$J = \frac{\partial}{\partial \theta}[E \quad \theta^T F^T] = [-K^T \quad F^T] \qquad (20)$$

$$E_{gdd} = [E \quad (F\theta)^T]^T \qquad (21)$$

The gradient algorithm – apart from these changes – remains the same. The difference is that here is no implicit inversion involved at the cost function level, therefore the computational needs are lower.

*3.5 Decrease the memory needs*

Section 3.3 and 3.4 address the inversion problems involved in the optimization and cost function evaluation. However, there is still a problem with the exaggerated memory usage which needs to be properly tackled.

In order to increase the computational performance, the Jacobian and gradient descent direction matrices are used in the form discussed in Section 3.2 ($J = [-K^T \quad R]$, $E_{gdd} = [E \quad \theta]^T$). The difference here is that a (directly defined) regularization filter matrix $R$ is used instead of $D$ which was computed via the inverted covariance matrix $P$. This step will allow us to gain some operational memory and computational time. In this case the first order DC and TC regularization filter functions are given by:

$$R_{1,DC}(i,j) = \begin{cases} \frac{1}{c_1} \frac{1+\rho^2}{\alpha^{(i+j)/2}(1-\rho^2)} & if\ i = j,\ i < n_\theta \\ \frac{1}{c_1} \frac{1+\rho^2}{\alpha^{(i+j)/2}(1-\rho^2)} & if\ i = j = 1\ or\ i = j = n_\theta \\ -\frac{1}{c_1} \frac{\rho}{\alpha^i(1-\rho^2)} & if\ |i-j| = 1 \\ 0 & otherwise \end{cases} \qquad (22)$$

$$R_{1,TC}(i,j) = \begin{cases} \frac{1}{c_1} \frac{1+\alpha}{\alpha^{(i+j)/2}(1-\alpha)} & if\ i = j,\ i < n_\theta \\ \frac{1}{c_1} \frac{1}{\alpha^{(i+j)/2}(1-\alpha)} & if\ i = j = 1\ or\ i = j = n_\theta \\ -\frac{1}{c_1} \frac{\sqrt{\alpha}}{\alpha^i(1-\alpha)} & if\ |i-j| = 1 \\ 0 & otherwise \end{cases} \qquad (23)$$

When analysing the observation and the regularization matrices one can clearly observe that they are highly structured. In fact, $K$ is a (lower) triangular Toeplitz matrix of the excitation signal and, the regularization matrices ($R, D, P$) are symmetric. These facts allow us to use special vectorial forms and instead of directly creating these matrices ($K, R$) let us define $K_{col}(n) \in \mathbb{R}^N$, $K_{row}(n) \in \mathbb{R}^{n_\theta}$, $R_{row}(n) \in \mathbb{R}^{n_\theta}$ as a function providing the n*th* column/row of the observation matrix $K$ and the regularization filter matrix $R$. In order to avoid misunderstanding, in the following part the multiplication between variables is explicitly marked by $\cdot$.

The algorithm presented in Section 3.2 remains the same, but certain elements are calculated element-wise/vector-wise.

1. The (vector) error term $E$ at $(k+1)$*th* iteration step is calculated element-wise (at each $n$ element) as follows:

$$E_{k+1}(n) = (y_{meas}(n) - K_{row}(n) \cdot \hat{\theta}_{k+1}),\ n = 0 \ldots N-1 \qquad (24)$$

2. The matrix product $J \cdot E_{gdd}$ is calculated at each $(k+1)$*th* iteration step in two steps, and it is stored in the vector $JE_{gdd} \in \mathbb{R}^{n_\theta}$. The first step provides the LS term of $J \cdot E_{gdd}$. It is calculated element-wise (at each $l$ element) as follows:

$$JE_{gdd_{k+1}}(l) = -K_{col}(l) \cdot E_{k+1},\ l = 0 \ldots n_\theta - 1 \qquad (25)$$

The second step provides the finial value of $J \cdot E_{gdd}$ as the sum of regularization term together with previously computed LS part (at each $l$ element). It is computed as follows:

$$JE_{gdd_{k+1}}(l) = JE_{gdd_{k+1}}(l) + R_{row}(l) \cdot E_{k+1},\ l = 0 \ldots n_\theta - 1 \qquad (26)$$

3. Using the previous term, the new (candidate) value of the estimator is calculated at each $(k+1)$*th* iteration step as:

$$\hat{\theta}_{k+1} = \hat{\theta}_k - \lambda \cdot JE_{gdd} \qquad (27)$$

4. Similarly to 2) the cost function $V$ at each $(k+1)$*th* iteration step is calculated in three steps. First, the LS term of the cost function is calculated:

$$V_{k+1} = E_{k+1}' \cdot E_{k+1} \qquad (28)$$

Second, the regularization term of the cost function calculated and stored in $\hat{\theta}R \in \mathbb{R}^{n_\theta}$ as follows:

$$\hat{\theta}R(l) = \hat{\theta}_{k+1}' \cdot R_{row}(l) \tag{29}$$

Finally, the cost function is given at $(k+1)th$ iteration step as follows:

$$V_{k+1} = V_{k+1} + gR \cdot \hat{\theta}_{k+1} \tag{30}$$

By using the proposed vectorised inversion-free gradient method the memory needs have been significantly decreased from $O(2N \times n_\theta + 3n_\theta \times n_\theta + N + n_\theta)$ to $O(N + 6n_\theta)$. In case of the illustrative example provided at the beginning of this section, the requested operational memory is decreased from 459 GB to 7 GB.

It is important to remark that, even though the memory needs are significantly smaller, the computational time will be much higher, due to the increased number of iterations.

*3.6 Practical considerations*

Due to the high computational needs of the memory reducing algorithm some practical advices are provided.

Whenever the memory needs are low ($N$ and/or $n_\theta$ is small, and consequently $K, D, P$ and $R$ are small) it is advised to use the simple inversion formula (see (4)) with filter based interpretation of the penalization term ($R$ matrix), instead of computing the inverse of the covariance matrices as it is the classical case. In this case – with no effort – multiple inversions of the covariance matrices are avoided resulting in significantly lower computing time.

For medium operational memory need, it is advised to use the algorithm presented in Section 3.4. If the memory needs are excessive then the algorithm in Section 435 is recommended. In both cases, the computational time can be shortened if in the first step, only the LS solution of the gradient based method is calculated (neglecting the regularization term), then using the LS estimate as the initial value to the regularized estimation, the whole regularized gradient algorithm is rerun.

A further important remark is that, despite the fact that it looks obvious to use a triangular decomposition (for instance, Chelovsky decomposition) in order to obtain the values of $F$ from $D$ or $R$, it is not recommended. The decomposition techniques provide an approximate solution with a relatively small error level. However, given the high number of parameters, these small errors accumulate and will quite likely result in divergence in the gradient method.

4. Transient Elimination

*4.1 Introduction*

When modelling dynamic systems, it is a common practice to use transient-free observations. This viewpoint is acceptable as long as a sufficiently long measurement is available. However, in many cases, this is not possible and without a proper transient elimination technique the estimated model can have a very low accuracy with respect to the underlying system. An excellent example can be, for instance, the benchmark problem of the cascaded water tanks: the measurements are very short and contain long transients. A further issue related to the benchmark problem is that the proposed nonparametric model is a (higher degree) Volterra model which requires an excessive number of model parameters (compared to the measurement length). Therefore, it is very important to clean up the measurement data by eliminating the undesired transient term.

In the case of nonparametric LTI system modelling, there are some common practices which can be used to eliminate the effect of the transient, for instance:

- If the excitation is periodic, then it is possible to discard some periods such that the measurement becomes transient free [30], [31]. The advantage of this technique is that an additional averaging can be used to reduce the effect of the measurement noise. The disadvantage is that at least some periods (and consequently a longer measurement) and a periodical input are needed which is not always possible.
- Different windowing methods can also help to eliminate the transient effects [38]. The disadvantage is that error level is still high, when a shorter measurement is used.
- There is also an alternative simple approach for nonparametric estimation [39]. This method uses a nonparametric IRF estimation technique. This can only be used if the transient term is very short compared to the steady-state IRF.
- A state-of-the-art nonparametric method is developed in [30], [31]. The Local Polynomial Method (LPM) is a local polynomial approximation of the smooth transfer function of the system and the smooth transient.

Unlike the LPM, the proposed method formulates the problem purely in the time-domain, and it makes use of some advanced prior information. This transient elimination method is based on [19] and [21], [26].

The interested reader can refer to [19], [29], [30] for detailed comparison of transient elimination techniques.

[21], [26] provide a detailed discussion on kernel based regularization in the context of transient free system identification and apply it to LTI systems. [19] discusses the possibilities of transient elimination for time-varying systems. The novelty of this work compared to these works is to formulate the regularization for the special case, when a short noisy measurement is disturbed by transient. This technique can not only be used as an extension of the proposed regularized Volterra kernel estimation but can be a handy tool for LTI identification and for nonlinear identification when the estimation problem can be formulated similarly.

*4.2 Assumptions*

It is important to remark that in the classical LTI theory the transient term is handled as a deterministic process. Unlike that, in the proposed technique it is seen from a Bayesian viewpoint: as the steady-state IRF, the transient IRF is also seen as a zero mean Gaussian random process [40] with smoothness and decaying properties. Using this perspective, we are allowed to use some advanced statistical methods but prior to carrying out any system identification procedure, several important assumptions must be addressed.

The first three assumptions are generally needed for the regularization technique. The last two assumptions are intended for the transient elimination technique. The last assumption is optional and its effect is studied later on.

*Assumption 1* The output is disturbed by additive, i.i.d. Gaussian noise with a zero mean and a finite variance, and the excitation signal is exactly known.

*Assumption 2* Each IRF, independently of its order and dimensionality, is smooth. This means that the spectral content of the underlying system is highly concentrated at low frequencies.

A more precise definition requires the theoretical details of the reproducing kernel Hilbert spaces, which is out of the scope of this article. A detailed description can be found in [40].

*Assumption 3* The observed system is bounded-input, bounded-output stable and the expected value of its steady-state and the transient IRFs are zero. Each IRF is shorter than the observation (measurement) length *(N)*.

In other words, this assumption implies that after a finite time, all IRFs tend to zero.

*Assumption 4* The transient IRF $h_{tr}(t)$ (and its vectorial form $\theta_{tr} \in \mathbb{R}^{n_{tr}}$) is independent of the noise, and its observable (measureable) length is $n_{tr}$.

*Assumption 5 (optional)* The past values of the excitation signal ($u_{past}$) are uncorrelated (or independent) of the future values ($u_{future}$).

Fulfilling this assumption simplifies the theoretical analysis and makes the practical implementation easier. The case when *Assumption 5* is not fulfilled is discussed later on.

Note, that the Gaussian white noise and the random phase multisine (the case of the cascaded water tanks benchmark) excitation signals automatically satisfy this assumption, because their samples are totally uncorrelated.

*4.3 The new baseline model*

In the basic Volterra baseline model it is assumed that the system is initially at rest, or in other words the initial conditions are zero. If this is not the case, transient errors will appear but only for a limited time and then they (exponentially) disappear. In this work this is seen via the following interpretation (see Fig. 4).

There is a finite observation window (with a length of $N$) wherein the measurement data are collected. Off course, we are aware of the fact that arbitrary excitation could have been applied before this window. Due to the insufficient knowledge of the past input values, we are unable to distinguish the contribution of the past and present input values in the first $n_{tr}$ samples (where the transient has an effect). If the observation is short, we need to use all the available data samples. In this work, instead of estimating or neglecting the past values of the input, we imagine the observable transient term (which appears in the first $n_{tr}$ samples) as an independent additive impulse response to a Kronecker delta function at $t = 0$ time instant (i.e. the excitation time of the first observed sample). An illustration is shown in Fig. 4.

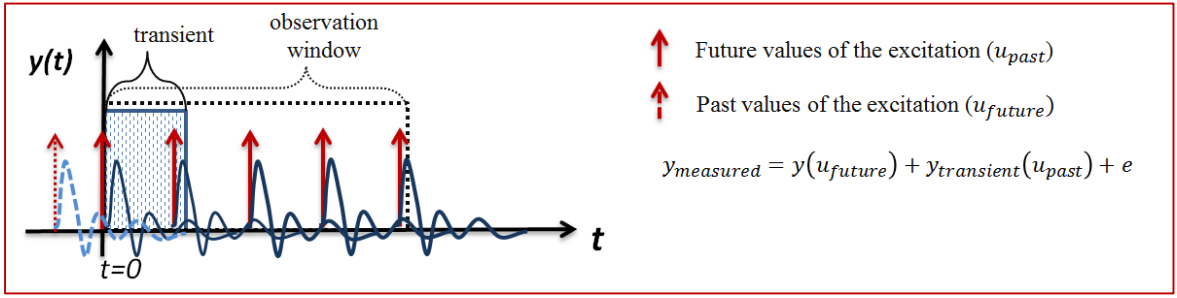

Fig 4. An illustration of the origin of the transient term. The red arrows refer to the excitation at different time instances. The black decaying curves refer to the impulse responses generated by the corresponding excitation. The dotted arrow and curve refer to the transient term. Observe that the transient impulse response function originates before the observation window and its observable (measurable) length is smaller than the whole IRF. If the past values of the excitation (i.e. the excitation before the observation window) are independent of the future values (i.e. the excitation in the observation window) i.e. *Assumption 5* is fulfilled, then the measured output and the transient response in the observation window are uncorrelated.

Using the assumptions given in Section 4.2 and the idea that the transient term can be imagined as an independent impulse response to a Kronecker delta function, (1) can be extended as follows:

$$y_{meas}(n) = \overbrace{h_0 + \sum_{m=1}^{M}(\sum_{\tau_1=0}^{n_m-1} \ldots \sum_{\tau_m=0}^{n_m-1} h_m(\tau_1,\ldots,\tau_m) \prod_{\tau=\tau_1}^{\tau_m} u(n-\tau))}^{y(n),\ Volterra\ term} + \overbrace{\sum_{\tau=0}^{n_{tr}-1} h_{tr}(\tau)\delta(n-\tau)}^{y_{tr}(n),\ transient\ term} + \overbrace{e(n)}^{noise} \quad (31)$$

where the impulse response function $h_{tr}(n) = 0$ and the transient contribution $y_{tr}(n) = 0$, when $n \geq n_{tr}$.

When *Assumption 5* is satisfied (see Fig. 4) then (31) boils down to:

$$y_{meas} = y(u_{future}) + y_{tr}(u_{past}) + e(n) \quad (32)$$

Equation (31) – (32) can be rewritten in vectorial form as ($K_\delta$ is explained later):

$$Y_{meas} = Y + Y_{tr} + E = K\theta + K_\delta \theta_{tr} + E \quad (33)$$

The new baseline model is illustrated in Fig 5.

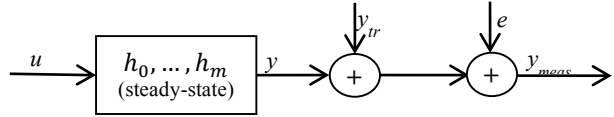

Fig 5. An illustration of the new baseline model in the presence of transient.

The key idea in the proposed technique is that we make use of the statistical framework, therefore a brief theoretical analysis is needed. The dependency – covariance – must be checked between the measured output and the transient term. Using this information, a modified estimator can be built. This dependency can be computed via the covariance:

$$\Sigma_{Y_{meas},Y_{tr}} = \mathbb{E}\{(Y_{meas} - \overbrace{\mathbb{E}\{Y_{meas}\}}^{0,\ As.\ 3})(Y_{tr} - \overbrace{\mathbb{E}\{Y_{tr}\}}^{0,\ As.\ 3})\} = \mathbb{E}\{(Y_{meas})(Y_{tr}^T)\} = \mathbb{E}\{(Y + Y_{tr} + E)(Y_{tr}^T)\} =$$

$$\mathbb{E}\{YY_{tr}^T\} + \mathbb{E}\{Y_{tr}Y_{tr}^T\} + \overbrace{\mathbb{E}\{EY_{tr}^T\}}^{0,\ As.\ 1,3,4} = \begin{cases} \mathbb{E}\{Y_{tr}Y_{tr}^T\}, & if\ As.\ 5 \\ \mathbb{E}\{YY_{tr}^T\} + \mathbb{E}\{Y_{tr}Y_{tr}^T\}, otherwise \end{cases} \quad (34)$$

This result will allow us to extend the presented regularization method by modifying the observation and penalization matrices. It can be observed, when *Assumption 5* is satisfied, that the transient term is uncorrelated (independent) of the true output which can simplify the proposed algorithm.

*4.4 The new cost function*

It is very important to remark that the transient IRF has inherently similar properties as the steady-state (higher order) FIR: it is smooth and exponentially decaying [30]. The aim of this section is to provide a methodology to estimate the transient IRF and to decrease its effects.

First, an extended parameter vector has to be defined. This is the joint parameter vector consisting of the steady-state Volterra and transient coefficients, and it is defined as $\theta' = [\theta^T\ \theta_{tr}^T]^T,\ \theta' \in \mathbb{R}^{n_\theta + n_{tr}}$.

Second, *Assumption 4* allows us to define the observation matrix of the transient response $K_\delta \in \mathbb{R}^{N \times n_{tr}}$ as a rectangular identity matrix.

From here the joint observation matrix is given by $K' = [K \ K_\delta]$, $K' \in \mathbb{R}^{N \times n_\theta + n_{tr}}$.

Next, the joint penalization matrix $D'$ needs to be defined. Its structure depends on the satisfaction of *Assumption 5*.

- Case 1: *Assumption 5* is not fulfilled

When the true output of the underlying system and the transient term are correlated, the joint penalization matrix $D'$ is given by:

$$D' = \begin{bmatrix} D & D_{cross} \\ D_{cross}^T & D_{tr} \end{bmatrix} \tag{35}$$

where $D' \in \mathbb{R}^{n_\theta + n_{tr} \times n_\theta + n_{tr}}$, $D_{cross} \in \mathbb{R}^{n_\theta \times n_{tr}}$ is the penalization matrix of the correlated elements and $D_{tr} \in \mathbb{R}^{n_{tr} \times n_{tr}}$ is the penalization matrix of the transient IRF and $D_{tr}$ needs to be defined as an ordinary LTI penalization matrix. $D_{cross}$ has to be defined similar way as $D$ is defined: when constructing $D_{cross}$ we have to use the cross terms of the corresponding IRF coefficients between $(h_0, \ldots, h_m)$ and $h_{tr}$ such that

$$D_{cross} = P_{cross}^+ \tag{36}$$

$$P_{cross} = \mathbb{E}\big[[\theta_0 \ \theta_1 \ \ldots \ \theta_m] [\theta_{tr}]^T\big] \tag{37}$$

where $P_{cross}$ is the covariance matrix between $(h_0, \ldots, h_m)$ and $h_{tr}$. $P_{cross}$ can have the structure of a TC/DC kernel (see Section 2.3), and the + matrix operator stands for the Moore-Penrose generalized pseudoinverse.

In practice, this method does not lead to a significantly better result and moreover, it would result to minimum two extra hyperparameters to tune (TC kernel). Under this condition, it is advised to consider the fulfilment of *Assumption 5* and use the following case:

- Case 2: *Assumption 5* is fulfilled

In the case when *Assumption 5* is fulfilled, the true output of the underlying system and the transient output are uncorrelated. In this case the joint penalization matrix boils down to:

$$D' = \begin{bmatrix} D & 0 \\ 0^T & D_{tr} \end{bmatrix} \tag{38}$$

where $D' \in \mathbb{R}^{n_\theta + n_{tr} \times n_\theta + n_{tr}}$ and $0 \in \mathbb{R}^{n_\theta \times n_{tr}}$ is a null matrix.

The above mentioned terms allow us to easily redefine the original cost function (3) as:

$$V' = \|Y_{meas} - K'\theta'\|_2^2 + \theta'^T D'\theta' \tag{39}$$

The regularized solution of $\theta'$ is now given by:

$$\hat{\theta}'_{reg} = \underset{\theta'}{\mathrm{argmin}} V' = (K'^T K' + D')^{-1} K'^T Y_{meas} \tag{40}$$

*4.5 Tuning the model complexity*

The choice of the transient model order ($n_{tr}$) and the hyperparameters values in $D_{tr}$ ($D_{cross}$) play a key role in the transient estimation procedure. Due to the smoothness and decaying assumptions of the transient IRF, the numerical values can be determined by TC or DC functions using a set of hyperparameters. Note that the possible kernel functions are not limited to TC/DC only and other kernel functions can be used as well, as long as they satisfy the assumptions made on the behaviour of the transient term.

For practical reasons we are interested in a simple model that describes the transient behaviour well. To determine an appropriate model order, several criteria can be used. A special residual analysis [29] is used here. In each validation step, the residuals are tested for whiteness and independency. When the dominating error is due to the noise, $E^T E$ gives a value around the level of the noise variance. A good model order is found when the squared difference gives a value around the noise variance and the residuals are white and independent. If, for different combinations of model order, the validation criterion is close to the variance level, the smallest order of the transient IRF is recommended to be used.

In the proposed implementation, the tuning of the hyperparameters is performed by the gradient descent method. When the obtained local minimum provides a large error, then the optimization method has to be repeated with a new set of initial values for the hyperparameters. The theoretical aspects of further possible optimization techniques are beyond the scope of this paper. Some relevant techniques can be found in [32].

*5. Volterra series estimation of the cascaded water tanks benchmark problem*

*5.1 Problem formulation*

In this section a brief overview of the benchmark problem is presented. The observed system and its measurements were provided, so the authors and other benchmark performers had no influence on the selection of the measurements. A detailed description of the underlying system, its measurements together with illustrative photos and videos can be found on the website of the workshop [41].

*5.1.1   Description of the cascaded water tanks*

The observed system consists of two vertically cascaded water tanks with free outlets fed by the (input) voltage controlled pump. The water is fed from a reservoir into the upper water tank which flows through a small opening into the lower water tank, where the water level is measured. Finally, the water flows through a small opening from the lower water tank back into the reservoir. The whole process is illustrated in Fig. 6.

Under normal operating conditions the water flows from the upper tank to the lower tank, and from the lower tank back into the reservoir. This kind of flow is weakly nonlinear [41] (see the square root terms in (41)–(42)). However, when the excitation signal is too large for certain duration of time, an overflow (saturation) can happen in the upper tank, and with a delay also in the lower tank. When the upper tank over flows, part of the water goes into the lower tank, the rest flows directly into the reservoir. This kind of saturation is strongly nonlinear. Without considering the saturation effect (overflow), the following input-output model can be constructed based on Bernoulli's principle and conservation of mass:

$$\dot{x}_1(n) = -k_1\sqrt{x_1(n)} + k_4 u(n) + w_1(n) \tag{41}$$

$$\dot{x}_2(n) = k_2\sqrt{x_1(n)} - k_3\sqrt{x_2(n)} + w_2(n) \tag{42}$$

$$y(n) = x_2(n) + e(n) \tag{43}$$

where $u(n)$ is the input (pump) signal, $x_1(n)$ and $x_2(n)$ are the states of the system (water tanks), $w_1(n)$, $w_2(n)$ and $e(n)$ are additive noise sources, and $k_1, k_2, k_3, k_4$ are constants depending on the system properties.

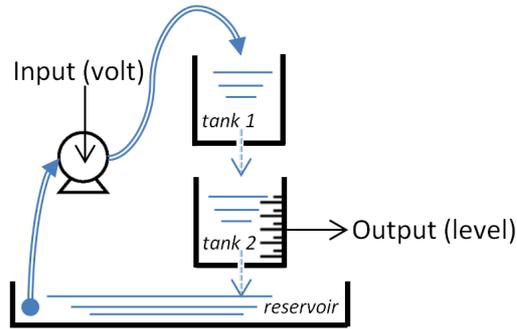

Fig 6. Schematic of the underlying system: the water is pumped from a reservoir in the uppertank, flows to the lower tank and finally flows back into the reservoir. The input is the pump voltage, the output is the water level of the lower tank.

*5.1.2   Description of measurement*

The excitation signal is a random phase multisine [30] with a length ($N$) of 1024. The sampling frequency is 0.25 Hz. The excited frequency ranges from 0 to 0.0144 Hz. The Signal-to-Noise Ratio (SNR) is around 40 dB. There are two datasets available for model estimation and validation, respectively. Each measurement starts in an unknown initial state. This unknown state is approximately the same for both the estimation and validation datasets. Furthermore, each dataset contains two overflows at two different time instances.

The water level is measured using uncalibrated capacitive water level sensors which are considered to be part of the system.

*5.1.3   The goal*

This work aims at obtaining a nonparametric time domain model based on the estimation data available. The goodness of the model fit is measured via the validation data on which the RMS error (see (44)) is calculated. The obtained error levels, together with an analysis of the user-friendly method will allow potential users to fairly compare different methods and will motivate them to use the proposed method.

The water tanks system has been modelled with the truncated time domain Volterra series (see (1)) of degree up to three. Moreover, for the purpose of completeness we include the identification results when the system is modelled as a FIR (only first order Volterra kernel, absence of zeroth kernel). In order to evaluate the performance of each model the following error index is used:

$$e_{RMSt} = \sqrt{1/N_v \sum_{n=1}^{N}(y_{mod}(n) - y_{val}(n))^2} \qquad (44)$$

where $y_{mod}$ is the modeled output and $y_{val}$ is the output provided in the validation data set.

The hyperparameter tuning for the Volterra series and the transient response estimation has been implemented with the Marginal Likelihood and the residual analysis method, respectively. For both cases the Volterra models have also been obtained with the inversion-free method (see Section 3), and all the results on the performance of the models as well as of the different algorithms are summarized in Table.1. In Fig. 7–9 block structures equivalent to the identified models are provided for illustration purposes ($G_0(q)$, $G_1(q)$ and $G_2(q)$ represent linear FIR blocks).

*5.2 Results*

*5.2.1 First order Volterra kernel*

The system is first modelled with a simple FIR model depicted in Fig. 7a. Even though the prior information of stability and smoothness is imposed on the modelled response as suggested by Fig. 7b, it is clear from Fig. 7c that the linear dynamics are not sufficient to describe the input - output behaviour of the water tanks system.

*5.2.2 First degree Volterra series*

The system is now modelled as a first degree Volterra series (FIR plus the constant zeroth Volterra kernel (i.e. offset), Fig. 8a). It can be observed again that the regularized first order Volterra kernel exhibits the prior information imposed during estimation (Fig. 8, right). The input-output behaviour of the underlying water tanks system is better described (Fig. 8c) compared to the previous case of FIR modelling, however it is clear that the extension of the series to higher degrees is necessary in order to capture the nonlinear behaviour of the system.

*5.2.3 Second and higher degree Volterra series*

The second degree Volterra series is used as a model for the true system (Fig. 9a). The regularized estimated second order Volterra kernel depicted in Fig 9b shows the effect of prior information compared to the unregularized estimated Volterra kernel on Fig. 9b (left).

The performance for the second degree Volterra series used as a model of the water tanks system is illustrated in Fig, 9c. The modelled output resembles more the true output compared to the previous cases of lower degree Volterra series (especially around the saturation), as expected. The results at this point indicate that the higher the degree of the series, the bigger the part of the nonlinear dynamics that will be described by the series.

It is important to highlight that despite the expectation, the 2D IRF shown in Fig. 9c. is not perfectly decaying. The reason is that a dominant decaying would require higher number of lags and thus, higher number of parameters. However, due to the limited number of measurement samples ($N = 1024$) and the excessive number of parameters ($n_\theta \approx 900$) a higher number of parameters would significantly increase the rms fitting error.

Third degree Volterra series model has also been identified. The error index for the third degree case can also be found in Table 1, the model structure and the validation outputs are shown in Fig. 10.

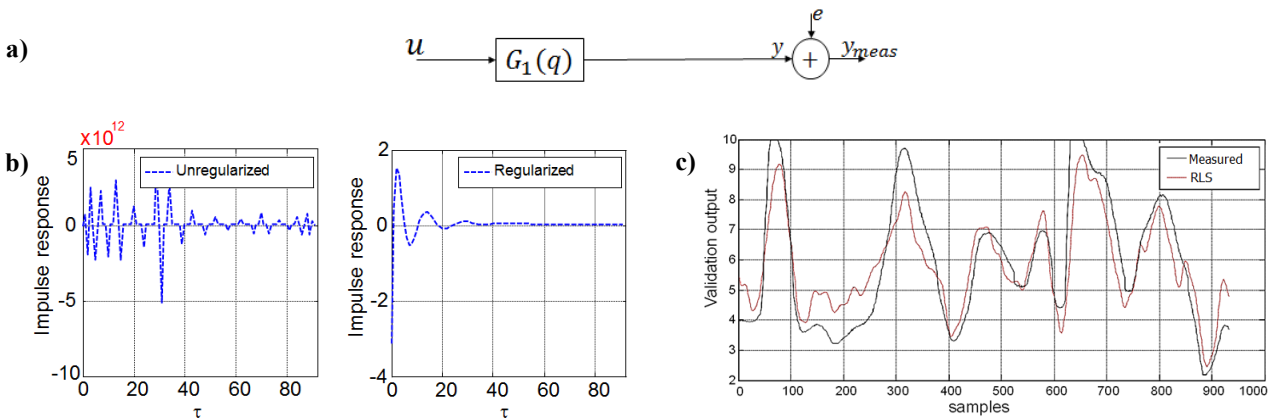

Fig 7. a) Block structure of the linear FIR model considered in the first case. b) Unregularized (left) and regularized (right) FIR estimation. c) True (black) and modelled (red) system output.

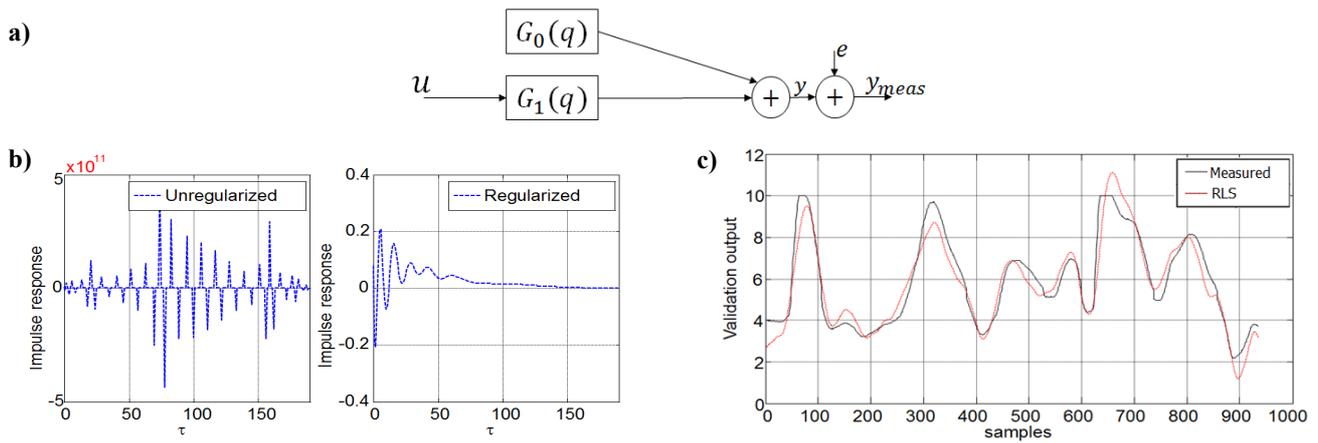

Fig 8. a) Block structure of the first degree Volterra series. b) Unregularized (left) and regularized (right) first order Volterra kernel estimation. c) True (black) and modelled (red) system output.

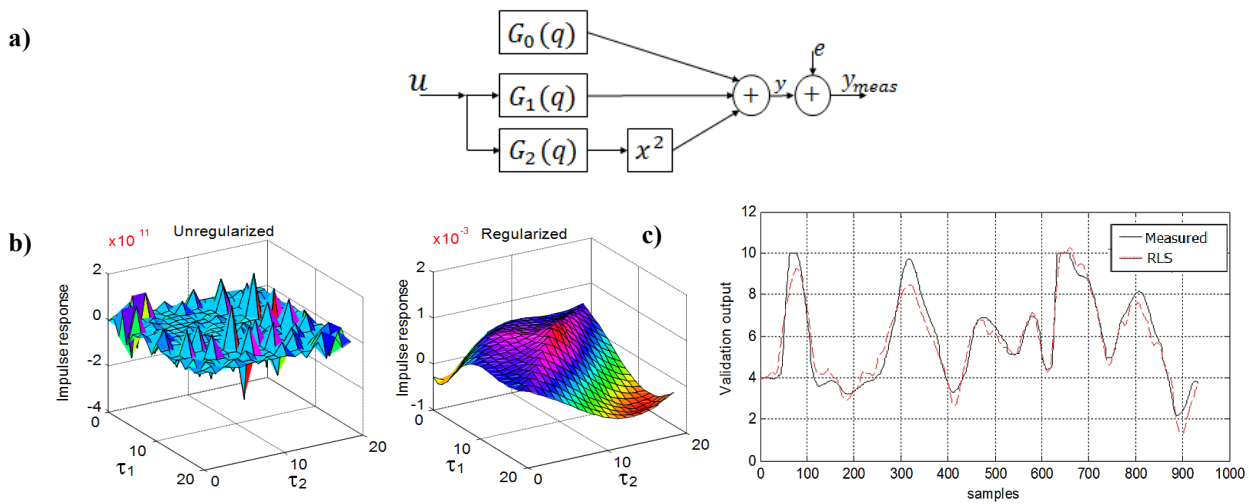

Fig 9. a) Block structure of the second degree Volterra series. b) Unregularized (left) and regularized (right) second order Volterra kernel estimation. c) True (black) and modelled (red) system output.

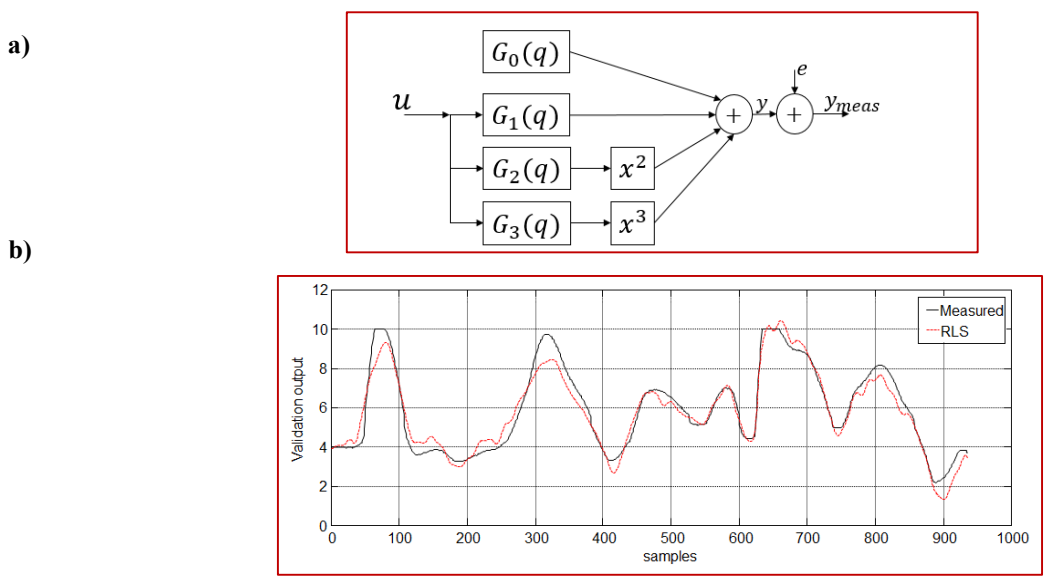

Fig 10.  a) Block structure of the third degree Volterra series. b) True (black) and modelled (red) system output.

*5.3 The effect of transients*

The model outputs shown in Fig. 7–10 are calculated on the transient-free estimation dataset. Figure 9c shows the second order model fit on the transient-free dataset, and Fig. 11 shows the model fit on the entire validation dataset. It can be observed that the level of transients is very high. The algorithm explained in Section 4 is used to remove the undesired transient effects (see Fig. 11, Table 1). In this particular case DC kernels were used. The hyperparameters are obtained by using a residual analysis where the residuals are tested for whiteness and independency. This is possible in this case because the transient appears on the quasi-linear part of the measurement (before the saturation). In this scenario, the residual analysis delivered a little bit better results than the empirical Bayes method (with respect to the error index).

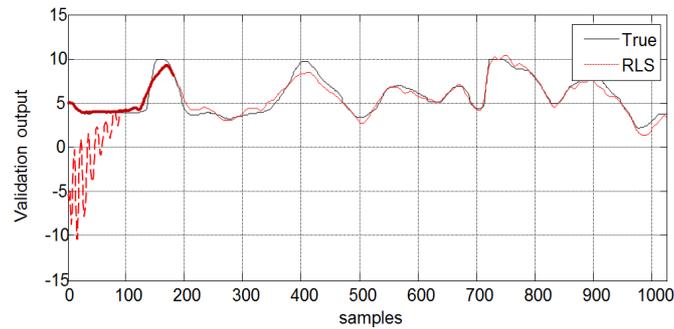

Fig 11. Entire validation output is shown for the 2nd degree Volterra model. Black line shows the true output. The red line shows the modelled output: dashed part is without transient estimation, and the solid part is with transient estimation.

*5.4 Summary*

The numerical results are given in Table 1. It can be clearly observed that the error level decreases as the degree of Volterra series increases. However, the error levels of the second and third degree models are quite close to each other. The latter is due to the fact that the third order Volterra kernel had only a limited number of parameters which is in turn due to the limited amount of measurement samples.

At this point we comment on the computational complexity which increases with increasing degree of the Volterra series (increasing number of parameters, see Fig. 4). In this particular situation, the time needed for a regularized FIR estimate is in the order of milliseconds. The first degree case has a computational time in the magnitude of seconds, the second degree Volterra kernel is in the magnitude of minutes, and finally the third degree Volterra series estimation requires hours. In Table 1 the total time needed for obtaining the best possible estimate is shown (including several estimated models for different memory lengths).

The same results have been obtained with the proposed inversion-free memory reduction method resulting in the same final numerical results. In that scenario, the requested operational memory has dropped to MB level from GB level. On the other hand, the computational time has been dramatically increased to days in case of third degree Volterra series. Of course, the exact computational time depends on several factors such as the model complexity, the initial values of the hyperparameters and the technique used to tune the hyperparameters.

**Table 1. The error indices for Volterra models of different degree are shown.**

| Method/ degree | Error index without transient estimation | Error index using transient removal | Computational time | | Memory needs | |
| --- | --- | --- | --- | --- | --- | --- |
| | | | with inversion | without inversion | with inversion | without inversion |
| FIR | 1.72 | 0.84 | ≈3 minutes* | | ≈4 MB* | |
| 1st degree | 1.56 | 0.59 *(-30%)* | ≈3 minutes* | | | |
| 2nd degree | 1.89 | 0.55 *(-35%)* | ≈3.5hours | ≈12hours | ≈80 MB | ≈0.1 MB |
| 3rd degree | 2.08 | 0.54 *(-36%)* | ≈6.5hours | ≈27hours | ≈850 MB | ≈0.8 MB |
| Extended white box model [42] | 0.28 | | - | - | - | - |

*only the classical inversion method has been executed

6. Conclusions

In this paper an efficient nonparametric identification method for nonlinear systems has been presented. Using the proposed methodologies, it is shown that the estimated Volterra series models of the cascaded water tanks benchmark system are able to capture the underlying dynamics, and their performance is comparable with the white-box (physical) models (see Table 1, [42]). Estimation of a model for this nonlinear benchmark system is challenging because:

- The number of samples is limited while a nonparametric modelling technique is used (large number of model parameters).
- The underlying system is weakly nonlinear but it exhibits a hard saturation effect for high peaks of the input signal.
- The measurement contains a relatively long transient effect.

In case of a weakly nonlinear system, a low order Volterra series is sufficient but in order to properly describe the saturation effect, an increased order of Volterra series should be used. Due to the limited number of samples and the excess of model complexity, the highest degree of the series is kept to three.

The regularization technique made it possible to achieve reliable results, even when the number of parameters was approximately equal to (or even higher than) the number of measurements. The computed error levels are comparable with best performing white-box (physical) model presented in the benchmark workshop [42] where the required time for model constructing is very high, i.e. it is very expensive.

In addition to these, the transient term could be eliminated with a simple extension of LTI regularization technique.

To conclude, it is shown that with an inexpensive general nonparametric modelling technique it is possible to achieve models of significant accuracy within a reasonable computation time.

**Acknowledgments**

This work was funded by the Methusalem grant of the Flemish Government (METH-1), the Fund for Scientific Research (FWO-Vlaanderen), and the Federal Government through the Inter university Poles of Attraction IAP VII/19 DYSCO program, by the ERC advanced grant SNLSID, under contract 320378, and by the VLAIO Innovation Mandate project number HBC.2016.0235**References**

[1] M. Schetzen, The Volterra and Wiener Theories of Nonlinear Systems, New York: Wiley, 1980.

[2] S. Boyd, L.O. Chua, "Fading Memory and the Problem of Approximating Nonlinear Operators with Volterra Series," *IEEE Trans. on Circuits and Systems,* vol. 32, no. 11, pp. 1150-1171, 1985.

[3] A. Stenger, L. Trautmann, R. Rabenstein, "Nonlinear acoustic echo cancellation with 2nd order adaptive Volterra filters," *1999 IEEE International Conference on Acoustics, Speech, and Signal Processing,* pp. 877-880, 1999.

[4] J-P. Costa, A. Lagrange, A. Arliaud, "Acoustic echo cancellation using nonlinear cascade filters," *2003 IEEE International Conference on Acoustics, Speech, and Signal Processing,* vol. 5, 2003.

[5] D. T. Westwick, R. E. Kearney, Identification of nonlinear physiological systems, New York: Wiley-IEEE Press, 2003.

[6] T. Koh, E. Powers, "Second-order Volterra filtering and its application to nonlinear system identification," *IEEE Trans. on Acoustics, Speech, and Signal Processing,* vol. 33, no. 6, pp. 1445-1455, 2003.

[7] T. Wu, A. Kareem, "Vortex-induced vibration of bridge decks: Volterra series-based model," *Journal of Engineering Mechanics,* vol. 139, no. 12, pp. 1831-1843, 2013.

[8] Y. Wan, T.J. Dodda, C.X. Wongb, R.F. Harrisona, K. Wordenb, "Kernel based modelling of friction dynamics," *Mechanical Systems and Signal Processing,* vol. 22, no. 1, pp. 66-80, 2008.

[9] B. Badji, E. Fenaux, M. E. Bagdouri, A. Miraoui, "Nonlinear single track model analysis using Volterra series approach," *Vehicle System Dynamics,* vol. 47, no. 1, pp. 81-98, 2009.

[10] H. Tang, Y.H. Liao, J.Y. Cao, H. Xie, "Fault diagnosis approach based on Volterra models," *Mechanical Systems and Signal Processing,* vol. 24, no. 4, pp. 1099-1113, 2010.

[11] M. Rébillat, H. Rafik and N. Mechbal, "Nonlinear structural damage detection based on cascade of Hammerstein models," *Mechanical Systems and Signal Processing,* vol. 48, no. 1, pp. 247-259, 2014.

[12] C.M. Cheng, Z.K. Peng, W.M. Zhang, G. Meng, "Volterra-series-based nonlinear system modeling and its engineering applications: A state-of-the-art review," *Mechanical Systems and Signal Processing,* Vols. 87, part A, pp. 340-364, 2017.

[13] P.S.C Heuberger, P.M. J Van den Hof, B. Wahlberg, Modelling and Identification with Rational Orthogonal Basis Func-, London, UK: Springer, 2005.

[14] A. da Rosa, R.J.G.B Campello, W.C. Amaral, "Choice of free parameters in expansions of discrete-time Volterra models using Kautz functions," *Automatica,* vol. 43, pp. 1084-1091, 2007.

[15] G. Favier, T. Bouilloc, "Parametric complexity reduction of Volterra models using tensor decompositions," in *17th European Signal Processing Con-*, Glasgow, Scotland, 2009.

[16] G. Pillonetto, F. Dinuzzo, T. Chen, G. De Nicolao, L. Ljung, "Kernel methods in system identification, machine learning and function estimation: A survey," *Automatica,* vol. 50, no. 3, pp. 657-682, March 2014.

[17] J. Lataire, T. Chen, "Transfer function and transient estimation by Gaussian process regression in the frequency domain," *Automatica,* vol. 72, p. 217–229, 2016.

[18] G. Birpoutsoukis, A. Marconato, J. Lataire J. Schoukens, "Regularized Nonparametric Volterra Kernel Estimation," *Automatica,* doi.org/10.1016/j.automatica.2017.04.014, 2017.

[19] P.Z. Csurcsia, Nonparametric identification of linear time-varying systems, Zelzate: Uitgeverij University Press, 2015.

[20] G. Birpoutsoukis, P. Z. Csurcsia, "Nonparametric Volterra series estimate of the cascaded tank," *Workshop on nonlinear system identification benchmarks,* p. 37, 2016.

[21] T. Chen, H. Ohlsson, L. Ljung, "On the estimation of transfer functions, regularizations and Gaussian processes—Revisited," *Automatica,* vol. 48, no. 8, pp. 1525-1535, June 2012.

[22] G. Pillonetto, M. H. Quang, A. Chiuso, "A new kernel-based approach for nonlinearsystem identification," *IEEE Transactions on Automatic Contro,* vol. 56, pp. 2825-2840, 2011.

[23] P. Z. Csurcsia, J. Lataire, "Nonparametric Estimation of Time-variant Systems Using 2D Regularization," *IEEE Transactions on Instrumentation & Measurement,* vol. 65, no. 5, pp. 1259-1270, May 2016.

[24] G. Pillonetto, A. Chiuso, G. De Nicolao, "Regularized estimation of sums of exponentials in spaced generated by stable spline kernels," in *American Control Conference*, Baltimore, 2010.

[25] D. K. Duvenaud, Automatic Model Construction with Gaussian Processes (PhD. thesis), Cambridge: Pembroke College, University of Cambridge, 2014.

[26] G. Pillonetto, G. De Nicolao, "A new kernel-based approach for linear system identification," *Automatica,* vol. 46, no. 1, pp. 81-93, January 2010.

[27] B. P. Carlin, T. A. Louis, Bayesian Methods for Data Analysis, 3rd ed., CRC Press, ISBN: 9781584886983, 2008.

[28] G. Pillonetto, A. Chiuso, "Tuning complexity in kernel-based linear system identification: The robustness of the marginal likelihood estimator," *2014 IEEE European Control Conference,* 2014.

[29] L. Ljung, System identification: Theory for the User, 2nd ed., New Jersey: Prentice-Hall, ISBN: 9780136566953, 1999.

[30] R. Pintelon, J. Schoukens, System Identification: A Frequency Domain Approach, 2nd ed., New Jersey: Wiley-IEEE Press, ISBN: 978-0470640371, 2012.

[31] J. Schoukens, R. Pintelon, Y. Rolain, Mastering System Identification in 100 exercises, New Jersey: John Wiley & Sons, ISBN: 978047093698, 2012.

[32] T. Chen, L. Ljung, "Implementation of algorithms for tuning parameters in regularized least squares problems in

system identification," *Automatica,* vol. 49, no. 7, pp. 2213-2220, July 2013.

[33] L. Ljung, "Model validation and model error modeling," in *Åström symposium on control*, Lund, Sweden, August 1999.

[34] L. Ljung, H. Hjalmarsson, "System identification through the eyes of model validation," *European Control Conferences,* pp. 940-954, 1993.

[35] J. Nocedal, S. J. Wright, Numerical Optimization (2nd ed.), Springer, 2006.

[36] A. Marconato, M. Schoukens, J. Schoukens, " Filter-based regularisation for impulse response modelling," *IET Control Theory & Applications,* October 2016.

[37] C. Van Loan, Introduction to scientific computing, 2nd edition, Prentice-Hall, 2000.

[38] G. Monteyne, Identification in Nuclear and Thermal Energy, Moderator Temperature Coefficient Estimation via Noise. (PhD thesis), Zelzate: Uitgeverij University Press, 2013.

[39] P. Z. Csurcsia, J. Schoukens, I. Kollár, "A first study of using B-splines in nonparametric system identification," in *IEEE 8th International Symposium on Intelligent Signal Processing*, Funchal, Portugal, 2013.

[40] C. E. Rasmussen, C. K. I. Williams, Gaussian Processes for Machine Learning, The MIT Press, ISBN: 0-262-18253-X, 2006.

[41] M. Schoukens, P. Mattson, T. Wigren, J.P. Noël, "Cascaded tanks benchmark combining soft and hard nonlinearities," Workshop on Nonlinear System Identification Benchmarks, Pp. 20–23, Brussels, URL: http://homepages.vub.ac.be/mschouke/ benchmark, 2016.

[42] G. Holmes, T. Rogers, E.J. Cross, N. Dervilis, G. Manson, R.J. Barthorpe, K. Worden, "Cascaded Tanks Benchmark: Parametric and Nonparametric," *Workshop on nonlinear system identification benchmarks,* p. 28, 25-27 April 2016.

[43] G. Pillonetto, A. Aravkin, "A new kernel-based approach for identification of time-varying linear systems," in *IEEE International Workshop on Machine Learning for Signal Processing (MLSP)*, Reims, 2014.

[44] A. H. Tan, K.R. Godfrey, H.A. Barker, "Design of ternary signals for MIMO identification in the presence of noise and nonlinear distortion," *IEEE Transactions On Control Systems Technology,* vol. 17, no. 4, pp. 926-933, Julx 2009.

[45] H.A. Barker, A.H. Tan, K.R. Godfrey, "Ternary input signal design for system identification," *IET Control Theory and Applications,* vol. 1, no. 5, pp. 1224-1233, September 2007.

[46] K.R. Godfrey, A.H. Tan, H.A. Barker, B. Chong, "A survey of readily accessible perturbation signals for system identification in the frequency domain," *Control Engineering Practice,* vol. 13, no. 11, pp. 1391-1402, November 2005.

[47] L. Ljung, "System identification toolbox," 1988-2012.

[48] Q. Rentmeesters, P.-A Absil, and Paul Van Doore, "Identification method for time-varying arx models," in *Recent Advances in Optimization and its Applications in Engineering*, Springer, ISBN: 9783642125980, 2010, pp. 193-202.